\documentclass[usenatbib]{mnras}

\usepackage[T1]{fontenc}
\usepackage{ae,aecompl}

\usepackage{amsmath}

\usepackage{graphicx}	
\usepackage{bm} 

\newcommand{\los}{\ensuremath{\mathrm{los}}}
\newcommand{\kms}{\ensuremath{\,\mathrm{km}\,\mathrm{s}^{-1}}}
\newcommand{\kpc}{\,\mathrm{kpc}}
\def\d{{\rm d}}
\def\vJ{{\bm J}}
\def\cR{\mathcal{R}}

\title[The Milky Way's globular cluster system]{Modelling the Milky Way's
globular cluster system}

\author[ J. Binney and L.K. Wong]{
James Binney$^{1}$\thanks{E-mail: binney@thphys.ox.ac.uk} and Leong Khim Wong$^{1,2}$
\\
$^{1}$Rudolf Peierls Centre for Theoretical Physics, Keble Road, Oxford OX1
3NP, UK\\
$^{2}$DAMTP, Centre for Mathematical Sciences, Wilberforce Road, Cambridge
CB3 0WA, UK\\
}

\date{Accepted XXX. Received YYY; in original form ZZZ}

\pubyear{2017}

\begin{document}
\label{firstpage}
\pagerange{\pageref{firstpage}--\pageref{lastpage}}
\maketitle

\begin{abstract}
We construct a model for the Galactic globular cluster system based on a
realistic gravitational potential and a distribution function (DF) analytic
in the action integrals. The DF comprises disc and halo components whose
functional forms resemble those recently used to describe the stellar discs
and stellar halo. We determine the posterior distribution of our model
parameters using a Bayesian approach. This gives us an understanding of how
well the globular cluster data constrain our model.  The favoured parameter
values of the disc and halo DFs are similar to values previously obtained
from fits to the stellar disc and halo, although the cluster halo system
shows clearer rotation than does the stellar halo.  Our model reproduces the
generic features of the globular cluster system, namely the density profile,
the mean rotation velocity. The fraction of disc clusters coincides with
the observed fraction of metal-rich clusters.  However, the
data indicate either incompatibility between catalogued cluster distances and
current estimates of distance to the Galactic Centre, or failure to identify
clusters behind the bulge.  As the data for our Galaxy's components increase
in volume and precision over the next few years, it will be rewarding to
revisit the present analysis.
\end{abstract}




\section{Introduction}
\label{sec:Introduction}

In recent years, it has become possible to construct sophisticated models for
the Milky Way. These models assume a gravitational potential that is
axisymmetric, and components of the Galaxy are approximated by distribution
functions (DFs) chosen to be analytic in three isolating integrals of motion.
This allows for the construction of equilibrium DFs via Jeans'
(\citeyear{JeansTheorem}) theorem. Candidate DFs for the stellar discs were
proposed by \cite{BinneyModels} and refined and extended by
\cite{BinneyMcMillan2011} and \cite{SandersBinney}. A DF for the dark halo
was implemented by \cite*{DarkHalo}, while \cite{Posti} and
\cite{WilliamsEvans} describe a wide range of DFs for spheroidal systems,
some of which have been applied and extended by \cite{DasBinney} and
\cite{DasBW}. In this paper, we explore how this technique of constructing
action-based DFs can be used to understand the Galactic globular cluster (GC)
system.

\cite{HarrisCanterna1979} noted that the distribution in metallicity of
the Galactic globular clusters is bimodal.  \cite{Zinn1985} showed that the
metal-rich and metal-poor sub-populations have distinct phase-space
distributions, the metal-rich clusters being more strongly concentrated to
the Galactic centre and forming a more rapidly rotating body.  Now one
divides the 157 GCs in the
\citeauthor{HarrisCatalog}~\citeyear{HarrisCatalog} catalogue (2010
edition)\footnote{\url{http://www.physics.mcmaster.ca/Globular.html}}
into 44 metal-rich GCs
with $\hbox{[Fe/H]}> -0.8$ and 113 metal-poor GCs with $\hbox{[Fe/H]}<-0.8$.
\cite{Zinn1993} later suggested that the metal-poor subpopulation could be
further divided into `young' and `old' based on the relation between their
metallicities and the morphologies of their horizontal branches \citep[see
e.g. Fig. 6 of][]{MackeyGilmore}. Moreover, he argued that the old
metal-poor GCs are more concentrated
closer to the Galactic centre than the young GCs. However, very precise ages
of 55 GCs extracted from HST data do not confirm a relationship between age
and Galactocentric radius \citep{vandenbergh2013}.
\cite{MackeyGilmore} found that the metal-rich GCs in
their sample were all within a Galactocentric radius $r \sim
6\,\mathrm{kpc}$, with 60 per cent of them situated at $r <
3.0\,\mathrm{kpc}$. Unlike the metal-poor GCs, this system is very flattened,
with all metal-rich GCs at $|z| < 3.6\,\mathrm{kpc}$, and all but four
located at $|z| < 2.0\,\mathrm{kpc}$ ($z$ is the vertical distance from the
plane of the Galaxy). 

Studying the velocity distribution of the GC system proves to be more
difficult, as 53 per cent of GCs do not have any proper motion data.
\cite{FrenkWhite} provide a method to estimate the mean
rotation velocity $v_\textup{rot}$ of the GC system based only on position
and line-of-sight velocities. Using this method, and assuming a circular
speed of $220\kms$ at the solar neighbourhood, \cite{Thomas1989} found that
the GC system has a mean rotation of $v_\textup{rot} \sim 65\kms$. The
metal-rich GCs in his sample systemically rotate with $v_\textup{rot} \sim
113\kms$, while the metal-poor GCs have $v_\textup{rot} \sim 43\kms$.
\cite{Zinn1985} expands on this result, finding that the old metal-poor GCs
rotate with $v_\textup{rot} \sim 70\kms$, while the younger metal-poor GCs
have a mean rotation consistent with zero, although the uncertainty in the
latter result is large.

These considerations are all consistent with Zinn's (\citeyear{Zinn1985})
identification that the metal-rich GCs exhibit disc-like kinematics, while
the metal-poor GCs exhibit halo-like kinematics. As the GC system is thought
to consist of these two distinct subpopulations, our DF will also be
constructed using distinct disc and halo components. However, in fitting the
DF to the data we make no assumption regarding the metallicities of the
components, and in this way investigate whether the GC system can be effectively
divided using phase-space data alone.

Our DF is described in Section~\ref{sec:Model}.  Section~\ref{sec:Inference}
explains how the posterior distribution for our model parameters was
determined.  Section~\ref{sec:Results} describes models favoured by the data,
including the spatial and kinematic properties of the two components.
Section~\ref{sec:Conclusions} sums up and suggests some directions for
further work.

\section{Model}
\label{sec:Model}

\subsection{The Galactic potential} \label{sec:GalacticPotential} 

\cite{PifflRave} sought an axisymmetric Galactic potential that is consistent
with a wide range of observational data. Specifically, they required the
potential to reproduce gas terminal velocities at various longitudes, the
kinematics of stellar masers with very precise astrometry, the proper motion
of Sgr~A*, the run of stellar density with distance from the Galactic plane
near the Sun, and the kinematics of $\sim 200\,000$ stars in the RAdial
Velocity Experiment \citep[RAVE;][]{RaveFirstRelease,RaveFourthRelease}.
The potential assumes that the Galaxy's mass is dominated by a gas disc, thin
and thick stellar discs, a stellar bulge and a dark halo. The disc components
are described by density distributions of the form
\begin{equation}
\rho_\textup{disc}(R,z) = \frac{\Sigma_0}{2 z_\mathrm{d}} 
\exp \left( - \frac{R}{R_\mathrm{d}} - \frac{R_\textup{hole}}{R} - \frac{|z|}{z_\textup{d}}\right),
\label{eq:DiscDensityProfile}
\end{equation}
where $(R,z)$ are cylindrical coordinates in the Galactocentric frame,
$R_\mathrm{d}$ and $z_\mathrm{d}$ are characteristic scales, and $\Sigma_0$
is the surface density of the disc. A non-zero value of $R_\textup{hole}$
allows for a central cavity.

The stellar bulge and dark halo have
\begin{equation}
\rho_\textup{halo}(R,z) = \frac{\rho_0}{m^a (1+m)^{b-a}} \exp \left[ -(m r_0 / r_\textup{cut})^2\right],
\label{eq:HaloDensityProfile}
\end{equation}
where
\begin{equation}
m(R,z) = \sqrt{(R/r_0)^2 + (z/q r_0)^2}.
\end{equation}
Here, $\rho_0$ is a normalisation constant, $r_0$ is a scale radius, and $q$
is the axis ratio for surfaces of constant density. The exponential term
permits the spheroids to extend only to a finite distance set by
$r_\textup{cut}$, and also ensures that the density profile has a finite mass
for all sensible values of $a$ and $b$. The Galactic potential $\Phi(R,z)$ is
then given by solving the Newton--Poisson equation.

As the constraints adopted by \cite{PifflRave} are probes of the vertical
profile of gravitating matter, both baryonic and dark, their data are
consistent with a model with a spherical dark halo and a heavier baryonic
disc, or a more flattened dark halo and a lighter baryonic disc.
Consequently, the axis ratio $q\leq 1$ of the dark halo is not tightly
constrained, although \cite{PifflRave} argue that a comparison with the
results of \cite{BienaymeEtAl} favours an axis ratio $q \simeq 0.8$. We have
chosen to use $q = 0.8$ for the dark halo, although a larger value has no
significant effect on our results. The full set of parameter values we used
for the gravitational potential are shown in
Table~\ref{table:PifflParameters}.  Section 2.3 of \cite{BinneyMcMillan2016}
gives details of downloadable code that evaluates $\Phi(R,z)$ and its
derivatives  given the numbers in Table~\ref{table:PifflParameters}.

\begin{table}
\caption{Parameters of our gravitational potential, which are fixed in our dynamical model.} 
\label{table:PifflParameters}
\centering
\begin{tabular}{lllll}
\hline
$\quad$         &
Thick disc $\;$ &
Thin disc  $\;$ &
Gas disc   $\;$ &\\
\hline
$\Sigma_0$        & 274.5 & 532.4 & 87.3 & $M_\odot\,\textup{pc}^{-2}$\\
$R_\textup{d}$    & 2.58  & 2.58  & 5.16 & kpc\\
$z_\textup{d}$    & 0.67  & 0.20  & 0.04 & kpc\\
$R_\textup{hole}$ & 0     & 0     & 4    & kpc\\
\hline
$\quad$            &
Stellar bulge $\;$ &
Dark halo     $\;$ &\\
\hline
$\rho_0$         & 94.9  & 0.0196 & $M_\odot\,\textup{pc}^{-3}$\\
$r_0$            & 0.075 & 15.5   & kpc                        \\
$r_\textup{cut}$ & 2.1   & 0      & kpc                        \\
$a$              & 0     & 1                                   \\
$b$              & 1.8   & 3                                   \\
$q$              & 0.5   & 0.8                                 \\
\hline
\end{tabular}
\end{table}

Note that the GC system does not enter into this model because its mass is
negligible. We treat the GCs as a system of
157 identical, non-interacting point particles moving in the static potential
$\Phi(R,z)$, each with Hamiltonian
\begin{equation}
H = {\textstyle\frac{1}{2}} \left( v_R^2 + J_\phi^2/R^2 + v_z^2\right) + \Phi(R,z).
\end{equation}
The velocity components $v_R$ and $v_z$ are in the radial and vertical
directions respectively, and $J_\phi \equiv R v_\phi$ is the conserved angular
momentum about the axis of Galactic rotation. Positive velocity $v_\phi$ is
in the direction of Galactic rotation.

\subsection{Angle--action variables}
We define the set of observables for a GC as
\begin{equation}
\bm u \equiv (l,b,s,v_\los,\mu_\alpha^*,\mu_\delta),
\end{equation}
where $(l,b)$ are the Galactic longitude and latitude, $s$ is the
heliocentric distance, $v_\los$ is the line-of-sight velocity, and $\bm\mu
\equiv (\mu_\alpha^*=\dot l\cos b,\mu_\delta=\dot b)$ is the proper motion vector. 

In order to transform these into the phase space coordinates
\begin{equation}
(\bm{x},\bm{v}) \equiv (R,\phi,z,v_R,v_\phi,v_z),
\end{equation}
we have assumed that the Sun is located at $(R,\phi,z) = (R_0,0,0)$, where
we take $R_0 = 8.3\,\mathrm{kpc}$ \citep{SchonrichSolarRadius}. We further
assume that the local circular speed is $240\kms$
\citep{SchonrichSolarRadius}, and that the Sun has velocity $\bm
(U,V,W)_{\odot} = (11.1,12.24,7.25)\kms$ \citep*{SunWrtLSR} relative to the
local standard of rest (LSR). Positive velocities $U$, $V$ and $W$ point in
the direction of the Galactic centre, Galactic rotation and Galactic north
pole respectively.

It proves useful to transform these phase space coordinates into
angle--action variables $(\bm{\theta},\bm{J})$. These are a set of canonical
coordinates where the momenta $J_i$ are integrals of motion. It then follows
from Hamilton's equations that $H \equiv H(\bm J)$ and the angles $\theta_i$
increase linearly with time \citep{BT08}.  The orbit of a GC is specified by
$\bm J$, and if at time $t = 0$ it is at position $\bm x(\bm\theta(0),\bm
J)$, then we simply increase the angles linearly to evolve the position
forwards in time. Despite its advantages, this formalism has until recently
been little used on account of the difficulty in evaluating $(\bm \theta,\bm
J)$. Recent technical progress now makes this possible. We make extensive use
of the St\"ackel Fudge \citep{BinneyActions,SandersBinneyRev}, which yields
$(\bm \theta,\bm J)$ given $(\bm x, \bm v)$ in an
axisymmetric potential like that of Section~\ref{sec:GalacticPotential}. We
identify $\bm J \equiv (J_r, J_z, J_\phi)$, where $J_r, J_z \geq 0$ can be
thought of as quantifying oscillations in the radial and vertical
directions.

\subsection{Distribution function}

Given that the GC system is in dynamical equilibrium, Jeans'
\citeyearpar{JeansTheorem} theorem allows us to assume that the DF is a
function $f(\bm J)$, so the probability that a randomly chosen GC has
phase-space coordinates in $\mathrm{d}^3\bm\theta\,\mathrm{d}^3\bm J$ is
\begin{equation}
\label{eq:DFSpaceEquality}
f(\bm J) \; \mathrm{d}^3\bm\theta\,\mathrm{d}^3\bm J  = f(J(\bm x, \bm v))\;
\mathrm{d}^3\bm x \, \mathrm{d}^3\bm v.
\end{equation}
We normalize $f(\bm J)$ such that
\begin{equation}
\label{eq:DFNormalization}
(2\pi)^3 \int \mathrm{d}^3\bm J \; f(\bm J) = 1,
\end{equation}
so the quantity $(2\pi)^3 f(\bm J)\;\mathrm{d}^3\bm J$ gives the probability
that a randomly selected GC moves on the orbit specified by $\bm J$.

{Although we do not assume that the metal-rich clusters form a
disc-like component and the metal-poor clusters form a spheroid, our DF is a
linear combination of a DF for a disc-like population and a DF for a halo
population.  We write
\begin{equation}
f(\bm J|\Pi) = F_\textup{disc} f_\textup{disc}(\bm J) + (1-F_\textup{disc})
f_\textup{halo}(\bm J), 
\end{equation}
where $F_\textup{disc}\in [0,1]$ is the fraction of disc GCs and
$\Pi$ is a set of parameters for the model. We will often write $f(\bm
J|\Pi)$ as simply $f(\bm J)$ in what follows, but they will mean the same
thing. It  turns out that favoured models assign  roughly as much probability
to the disc-like component as the fraction of observed clusters that are
metal rich, but this
is an empirical result rather than an assumption.}

The DF of  an axisymmetric system is usually best considered to be the sum of a
part $f_+$ even in $J_\phi$ and a part $f_-$ odd in $J_\phi$. The latter does
not contribute to the density of the system, but is largely responsible for
the system's rotation. If we wish to avoid  discontinuities in the DF, $f_-$
must vanish with $J_\phi$. A convenient way to satisfy this condition and
obtain a non-negative and physically reasonable DF is to posit
\begin{equation}\label{eq:defk}
f_-(\vJ)=k\tanh(J_\phi/L)f_+(\vJ).
\end{equation}
 Here $-1\le k\le1$ is a constant that controls the sign and intensity of any
rotation, and $L$ is a constant that controls the steepness of the system's
central rotation curve. We have used this ansatz for the odd parts of the DFs
of both disc and halo.

\subsubsection{Disc}

The disc component is described by  the
`quasi-isothermal' DF introduced by
\cite{BinneyMcMillan2011}.  We take the part even in $J_\phi$ to be
\begin{equation}
f_\textup{disc+}(\bm J) = \frac{\Omega\nu \Sigma}{2\pi^2\kappa \sigma_r^2\sigma_z^2} 
 \exp\left( -\frac{\kappa J_r}{\sigma_r^2} -\frac{\nu J_z}{\sigma_z^2} \right)
\end{equation}
where
\begin{equation}
\Sigma(J_\phi) = \Sigma_0 \exp\left[
 -\frac{R_\textup{c}(J_\phi)}{\cR_\textup{d}} \right].
\end{equation}
$R_\textup{c}(J_\phi)$ is the radius of the circular orbit with angular momentum
$J_\phi$, $\Sigma(J_\phi)$ is approximately the surface density of the disc at that
radius, $\cR_\textup{d}$ is a characteristic scale length, and $\Sigma_0$ is a
normalisation constant that ensures that $(2\pi)^3\int\d^3{\bm J}\,f_{\rm
disc}=1$.
The functions $\Omega(J_\phi)$, $\kappa(J_\phi)$ and $\nu(J_\phi)$ are the circular,
radial and vertical epicycle frequencies of the potential evaluated at the
radius $R_\textup{c}(J_\phi)$.  Following \cite{BinneyMcMillan2011} we
let the velocity-dispersions parameters $\sigma_r$ and $\sigma_z$ vary with
$J_\phi$ as
\begin{equation}\label{eq:defs}
\sigma_i=\sigma_{i0}\exp\left[(R_0-R_\textup{c})\gamma\right],
\end{equation}
 where $\sigma_{i0}$ and $\gamma$ are free parameters. 

\subsubsection{Halo}

Our halo DF is inspired by \cite{Posti}. \cite{DasBinney} and \cite{DasBW} recently
extended DFs of this type to depend on [Fe/H] and age in addition to $\bm J$.
However, given the small number of halo clusters, we doubt the ability of the
data to constrain dependence on [Fe/H]. So we model the halo clusters with a
metallicity-blind and age-blind DF. 

The DFs of \cite{Posti} are constructed from a function $h({\bm J})$ that is a
homogeneous function of degree one in the $J_i$. The resulting stellar
system has a plausible structure near the $z$ axis only when the coefficients
in $h$ of $J_\phi$ and $J_z$ become equal as $J_\phi\to0$. We ensure
satisfaction of this condition in the simplest possible way, namely by making
$h$ a function of $|J_\phi|+J_z$. In fact we define $h$ to be
\begin{equation}
\label{eq:HaloDFh}
h(\bm J) = J_\textup{core} + J_r + \frac{\Omega}{\kappa}( |J_\phi|+J_z),
\end{equation}
 where the small constant $J_{\rm core}=10^{-3}\kpc\kms$ ensures that the DF
remains finite at $\bm J = 0$ at the cost of making $h$ not quite a
homogeneous function of the $J_i$. The epicycle frequencies are evaluated at
$R_\textup{c}(J_\textup{tot})$, where $J_\textup{tot} = J_r + J_z +
|J_\phi|$.  This choice is made to prevent the epicycle frequencies becoming
large for eccentric or highly inclined orbits (when $|J_\phi|$ is small)
\citep{DarkHalo}. 

For the halo clusters, the part of the  DF of  that is even in $J_\phi$ is
\begin{equation}
\label{eq:HaloDF}
\begin{split}
f_\textup{halo+}(\bm J) =& N_0{[1+J_0/h(\bm J)]^\alpha \over [1 +  h(\bm
 J)/J_0]^{\beta}}
\exp\!\left[-\left(\frac{h(\bm J)}{J_\textup{max}}\right)^4\!\right]
\end{split}
\end{equation}
with $N_0$ a normalisation constant chosen so $(2\pi)^3\int\d^3{\bm
J}\,f_\textup{halo}=1$.  The exponents $\alpha$ and $\beta$ in
equation (\ref{eq:HaloDF}) control the inner and outer power-law slopes of
the DF, and the two regimes are separated by a break scale $J_0$.
\cite{Posti} show that DFs with $(\alpha,\beta) = (2, 5)$ self-consistently
produce models with radial density profiles that closely approximate that of
\cite{Jaffe}, while DFs with $(\alpha,\beta) =(5/3,5)$ and $(5/3,3)$
self-consistently produce models with radial density profiles very similar to
those of \cite{Hernquist} and \cite*{NFW} models, respectively.
Unfortunately, since the GC system does not generate the potential that
confines it, the relations given by \cite{Posti} between $\alpha,\beta$ and
the slopes of the density profile in real space do not
apply.
Lastly,
the exponential term with $J_\textup{max}=10^6\kpc\kms$ ensures that the DF
has a finite mass for all choices of $\alpha$ and $\beta$; this has little
effect on the DF provided we choose $J_\textup{max}$ to be sufficiently
large.

\section{Bayesian inference}
\label{sec:Inference}

To determine the values and associated uncertainties of the model parameters
$\Pi$ that best fit the GC data, we compute the posterior distribution
\begin{equation}
\textup{Pr}(\Pi|\textup{Data}) \propto \textup{Pr}(\textup{Data}|\Pi) \times \textup{Pr}(\Pi),
\label{eq:Posterior}
\end{equation}
where $\textup{Pr}(\Pi)$ is an appropriately chosen prior and
$\textup{Pr}(\textup{Data}|\Pi) \equiv \mathcal L$ is the data's likelihood.
We follow the approach to the application of Bayesian inference to Galactic
structure described in \cite{MB13} (hereafter MB13).

\begin{figure}
\centering \includegraphics[width=0.8\hsize]{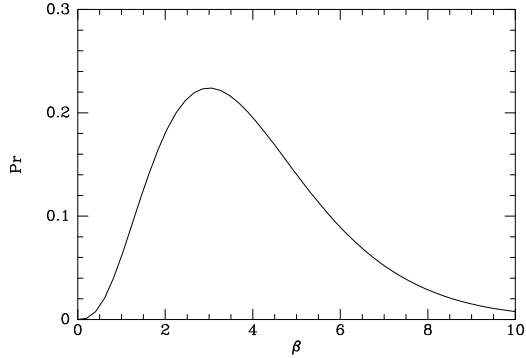}
 \caption{The adopted prior on the slope parameter $\beta$
 (eqn.~\ref{eq:PriorBeta}).} \label{fig:BetaPrior}
\end{figure}

\subsection{Prior distribution}
\label{sec:Prior}

As we are interested in how the GC data constrain the parameters of this
model, we opt for the least informative prior possible. The variables
$k_\textup{d},k_\textup{h} \in
[-1,1]$ and $F_\textup{disc} \in[0,1]$ have finite ranges, so we take the
prior to be uniform in these ranges.  The scale parameters $L_\textup{d}$,
$L_\textup{h}$, $J_0$, $\sigma_{i0}$,
and $\cR_\textup{d}$ are intrinsically positive, so the unbiased
prior is uniform in the logarithms of these quantities \citep{Jeffreys}. The
indices $\alpha,\beta$ can, in principle take arbitrarily large positive
values, but if we require only $\beta>0$, the data yield a degeneracy in
which both $\beta$ and $J_0$ increase.
From the work of \cite{Posti} and
\cite{DasBinney} we expect $\beta\sim5$.  To incorporate this knowledge
into our prior, we take the prior on $\beta$ to be the gamma distribution
\begin{equation}
\textup{Pr}(\beta) = \frac{B}{\Gamma(A)}(B\beta)^{A-1} \mathrm{e}^{-B\beta},
\label{eq:PriorBeta}
\end{equation}
where $A$ and $B$ are hyperparameters and $\Gamma(A)$ is the gamma function.
We set $A = 4$ and $B=1$ such that the distribution has a mean $\beta = 4$, a
mode $\beta = 3$, and a variance of $4$.  With this prior on $\beta$,
implausibly large values of $\beta$ and $J_0$ acquire low values of
the posteriori probability, but it might be argued that a weaker prior would
be preferable.  We have investigated two priors on the inverse
distance $\gamma$: uniform and uniform in its logarithm.

In summary, our prior is
\begin{equation}
\textup{Pr}(\Pi) \propto
\begin{cases}
 {\displaystyle\frac{\textup{Pr}(\beta)\,\textup{Pr}(\gamma)}{L_\textup{d} L_\textup{h} J_0
\sigma_r \sigma_z \cR_\textup{d}}}&
{\hsize=1.35in\raggedright\vbox{\noindent$k_\textup{d},k_\textup{h}\in[-1,1]$,
$F_\textup{disc}\in[0,1]$,
$L_i, \alpha, \beta, J_0, \sigma_{i0}, \cR_\textup{d} > 0$,}}  
\\
0 & \textup{otherwise,}
\end{cases}
\label{eq:Prior}
\end{equation}
where, as discussed in Section~\ref{sec:Results} $\textup{Pr}(\gamma)$ is
successively (i) uniform in $\gamma$, (ii) uniform in $\log\gamma$, and (iii) $\delta(\gamma)$.

\subsection{Data}

Data on the position variables $(l,b,s)$ of the 157 known GCs were obtained
from the 2010 edition of the \citeauthor{HarrisCatalog}~\citeyear{HarrisCatalog}
catalogue. This catalogue also
contains line-of-sight velocities for 143 of these GCs. Proper motion data
exist for only 64 GCs; these have been obtained from \cite{HSTProperMotion}
and a catalogue by \cite{CasettiDinescuLatest}\footnote{The proper motions
catalogue can be accessed at \url{http://www.astro.yale.edu/dana/gc.html}.}.

The Galactic coordinates $(l,b)$ are measured to high precision, so the
effects of their errors will be neglected. The errors in the remaining
observable quantities are assumed to be independent. If $u \in \{ v_\los,
\mu^*_\alpha, \mu_\delta \}$ denotes a component of velocity, then we
assume that its measured value $\bar u$ is related to its true value $u$ by
the Gaussian distribution
\begin{equation}
G(u,\bar u, \sigma_u) = \frac{1}{\sqrt{2\pi\sigma_u^2}} \exp \left[ - \frac{(u - \bar u)^2}{2\sigma_u^2} \right],
\end{equation}
where $\sigma_u$ is the uncertainty in $u$. Where possible, the values of
$\bar u$ and $\sigma_u$ are obtained from the data sources listed above. When
no such data are available, following MB13 the value of $\sigma_u$ is
taken to be sufficiently large that the Gaussian distribution is
effectively constant for all sensible values of $u$.

Since  heliocentric distance $s$ is an intrinsically positive quantity, we
take its  probability distribution to be a lognormal distribution
\begin{equation}
G_L(s,\bar s, \sigma_s) = \frac{1}{s} \frac{1}{\sqrt{2\pi\sigma_s^2}} \exp \left[ - \frac{(\ln s - \mathcal S)^2}{2\sigma_s^2} \right],
\end{equation}
where following \citep{CasettiDinescuLatest} the values of $\sigma_s$ and
$\mathcal S(\bar s, \sigma_s)$ are chosen such that the distribution has a
mean equal to the measured distance $\bar s$ and a variance of $(0.1\bar
s)^2$. Actually the uncertainties in cluster distances are surely sometimes
smaller and sometimes larger than this estimate, so this work could be
upgraded by obtaining errors on individual cluster distances. We shall see
also that it would be very desirable to take pains to ensure that all cluster
distances are assigned using the same value of $R_0$ as that adopted for
the model.

For brevity, we use the notation
\begin{equation}
\begin{split}
G(\bm u, \bar{\bm u}^\eta,\bm\sigma^\eta) =&\, \delta(l-\bar l^\eta) \times \delta(b - \bar b^\eta)
\\
&\times G_L(s,\bar s^\eta,\sigma_s^\eta) \times G(v_\los,\bar v_\los^\eta,\sigma^\eta_{v_\los})
\\
&\times G(\mu^*_\alpha, {\bar{\mu^*_\alpha}}^\eta,\sigma^\eta_{\mu^*_\alpha}) \times G(\mu_\delta,\bar\mu_\delta^\eta,\sigma^\eta_{\mu_\delta})
\label{eq:ErrorDist}
\end{split}
\end{equation}
to describe the six-dimensional error distribution for the observables of the
$\eta^\textup{th}$ GC.

\subsection{Likelihood}

While MB13 include the apparent magnitude $m$ as an observable, we do not
because we assume the completeness of our sample is independent of magnitude.
The likelihood $\mathcal L$ that the GC data
are drawn from the model $f(\bm J)$ is the product
\begin{equation}
\label{eq:Likelihood}
\mathcal L = \prod_\eta \mathcal L_\eta = \prod_\eta \int \mathrm{d}^6\bm{u}
\;  G(\bm u, \bar{\bm u}^\eta,\bm\sigma^\eta) \,\mathrm{Pr}(\bm u | \textup{Model}),
\end{equation}
where $\mathrm{Pr}(\bm u | \textup{Model})$ is the probability that a
randomly chosen GC has true observables $\bm u$:
\begin{equation}
\mathrm{Pr}(\bm u | \textup{Model}) = A \,S(\bm u)\,f(\bm J) \, 
\left| \frac{\partial(\bm\theta,\bm J)}{\partial(\bm u)}\right|,
\end{equation}
with
\begin{equation}
\left|\frac{\partial(\bm\theta,\bm J)}{\partial(\bm u)}\right| \equiv s^6 \cos b.
\end{equation}
Here the coordinates $(\bm\theta,\bm J)$ are evaluated at the given point $\bm u$,
and the selection function $S(\bm u)$ gives the probability that if a GC with
true observables $\bm u$ exists, it has been included in the Harris and
Casetti--Dinescu catalogues. The normalisation constant $A$ is given by the
condition $\int\d^6{\bm u}\,\mathrm{Pr}(\bm u | \textup{Model})=1$, so
\begin{equation}
1/A = \int \mathrm{d}^3\bm x\, \mathrm{d}^3\bm v \; 
f[\bm J(\bm x,\bm v)] S[\bm u(\bm x,\bm v)].
\label{eq:LikelihoodNorm}
\end{equation}

\subsubsection{Selection function}

Given that globular clusters have been discovered over many decades and in a
range of wavebands, it is impossible to characterise the incompleteness of
our sample with any precision.  It is, however, believed that nearly all our
Galaxy's GCs have been observed -- it is estimated that the total number of
Galactic GCs lies between 160 to 180 \citep{HarrisReview,Glimpse02}.  If this
is accepted, we do not need a sophisticated selection function. It is likely
that that any GCs that have still eluded astronomers are likely to lie close
to the Galactic plane, so they are hidden by dust. To take this effect into
account, we adopt the selection function
\begin{equation}
S(\bm u) =
\begin{cases}
1 & \textup{if } \left.\mathrm{E}(B-V)\right|_{(l,b)} < \mathrm{E}(B-V)_\textup{max} \\
0 & \textup{otherwise.}
\end{cases}
\end{equation}
This simply states that a GC located at Galactic coordinates $(l,b)$ will be
unobservable if the extinction in that direction is above a threshold value
$\mathrm{E(B-V)}_\textup{max}$.

We can obtain estimates for $\mathrm{E(B-V)}$ at any Galactic coordinate
using the dust map produced by \cite*{SFD}\footnote{Dust map values were
obtained from \url{http://irsa.ipac.caltech.edu/applications/DUST/}.}. We
find that we must choose $\mathrm{E(B-V)}_\textup{max} \geq 4$ to ensure that
all Harris catalogue GCs that were observed in the visible have $S(\bm u) =
1$. An important subtlety is that \cite{SFD} produce estimates for
extinctions from the observer to infinity, rather than from the observer to a
given heliocentric distance $s$. This is only an issue for GCs that might be
located in the Galactic plane but have sufficiently small values of $s$ that they
can nevertheless be detected. Such instances are probably rare we ignore them.

It turns out that a few GCs in the Harris catalogue have been discovered in
the infrared, and while this selection function does not include that
possibility, the number of such GCs is so small that the effect is
insignificant. Further, our results are not noticeably affected by any choice
of threshold value greater than $4$. In fact, excluding a selection function
all together does not dramatically alter our results. Below we keep
$\mathrm{E(B-V)}_\textup{max} = 4$.

Since we require values for $E(B-V)$ over the whole sky and it is essential
to use a consistent set of values, we do not use the values for individual
clusters given by Harris.

\subsubsection{Evaluating the likelihood}

We now use Monte Carlo methods to approximate the integrals in
equations~\eqref{eq:Likelihood} and \eqref{eq:LikelihoodNorm}. We introduce a
sampling density $f_\mathrm{S}(\bm x,\bm v)$ that approximates the phase
space distribution of the GC data. This ensures evaluations are concentrated
where the integrand is largest. The normalisation constant in
equation~\eqref{eq:LikelihoodNorm} can now be evaluated as
\begin{equation}
1/A \simeq \frac{1}{N_\textup{S}} \sum_{k=1}^{N_\textup{S}} 
\frac{f(\bm J(\bm x_k, \bm v_k))}{f_\textup{S}(\bm x_k,\bm v_k)} S[\bm u(\bm
x_k,\bm v_k)],
\label{eq:McLikelihoodNorm}
\end{equation}
where we draw $N_\textup{S}$ points $(\bm x_k,\bm v_k)$ from the sampling
density $f_\textup{S}$.  The integral in equation~\eqref{eq:Likelihood} for
the $\eta^\textup{th}$ star becomes
\begin{equation}
\mathcal L_\eta \simeq \frac{A}{N_\eta C_\eta} 
\sum_{k=1}^{N_\eta} \frac{f(\bm J(\bm x_k, \bm v_k))}{f_\textup{S}(\bm
x_k,\bm v_k)} S(\bm u),
\label{eq:McLikelihood}
\end{equation}
where we draw $N_\eta$ points $(\bm x_k,\bm v_k)$ from the sampling density
\begin{equation}
\xi(\bm u | \bar{\bm u}^\eta) = C_\eta G(\bm u, \bar{\bm
u}^\eta,\bm\sigma^\eta) f_\textup{S}(\bm x,\bm v)  
\left|\frac{\partial(\bm\theta,\bm J)}{\partial(\bm u)}\right|,
\end{equation}
The normalisation constant $C_\eta$ depends only on the data and not on the
model $f(\bm J)$, hence need not be computed in what follows.

\subsubsection{Choice of sampling density}
\label{sec:SamplingDensity}

A good choice for the sampling density $f_\textup{S}(\bm x,\bm v)$ is one
that approximates a typical model $f(\bm J(\bm x,\bm v))$. We have chosen
$f_\textup{S}$ to be a product of the density profile
$\rho_\textup{halo}(R,z)$ given in equation~\eqref{eq:HaloDensityProfile}
with $1/r_\textup{cut} = 0$ and a triaxial Gaussian velocity distribution
with principal axes aligned along the $v_R$, $v_z$ and $v_\phi$ directions.

A maximum likelihood fit to the GC data yields $a = 0$, $b = 4.43$, $r_0 =
2.49\,\mathrm{kpc}$ and $q = 0.83$ for the density profile. We have not
attempted to attach any confidence intervals to these numbers since all we
require is a crude first guess at the phase space distribution. For this same
reason, we have chosen to approximate the density profile using only
$\rho_\textup{halo}(R,z)$, rather than with an appropriate combination of
$\rho_\textup{halo}(R,z)$ and $\rho_\textup{disc}(R,z)$.

Choosing parameters for the velocity distribution of $f_\textup{S}$ is a more
nuanced task because only 64 of the 157 GCs have complete velocity data.
Naturally we take the means of $v_R$ and $v_z$ to be zero, while we set
$\langle v_\phi\rangle=v_\textup{rot}$, where $v_\textup{rot}=79\kms$ is the
rotation velocity returned by the algorithm of \cite{FrenkWhite} when applied
to our data set with the circular speed at the Sun set to
$v_\textup{c}=240\kms$. When the Frenk-White algorithm is applied to the
metal-rich subpopulation, we find $v_\textup{rot} = 137\kms$, while the
metal-poor population yields $v_\textup{rot}=48\kms$. These values will be
useful later. Note that these results are larger than those obtained by
\cite{Thomas1989} (see Section~\ref{sec:Introduction}). This is because we
have a larger data set and have also used a larger value for the local circular
speed.

As for the velocity dispersions of $f_\textup{S}$, we have chosen $(\sigma_R,
\sigma_\phi, \sigma_z) = (140,140,100)\kms$, based on the dispersions
calculated from the sample of 64 GCs with proper motion data. It is not
necessary that these dispersion values describe the GC system as a whole. To
ensure that we have made a reasonable choice, we have repeated our analysis
choosing different values of the dispersions in the range $100$--$250\kms$.
We find our results are not significantly affected for dispersions within
this range.

\begin{figure}
\centering
\includegraphics[width=\hsize]{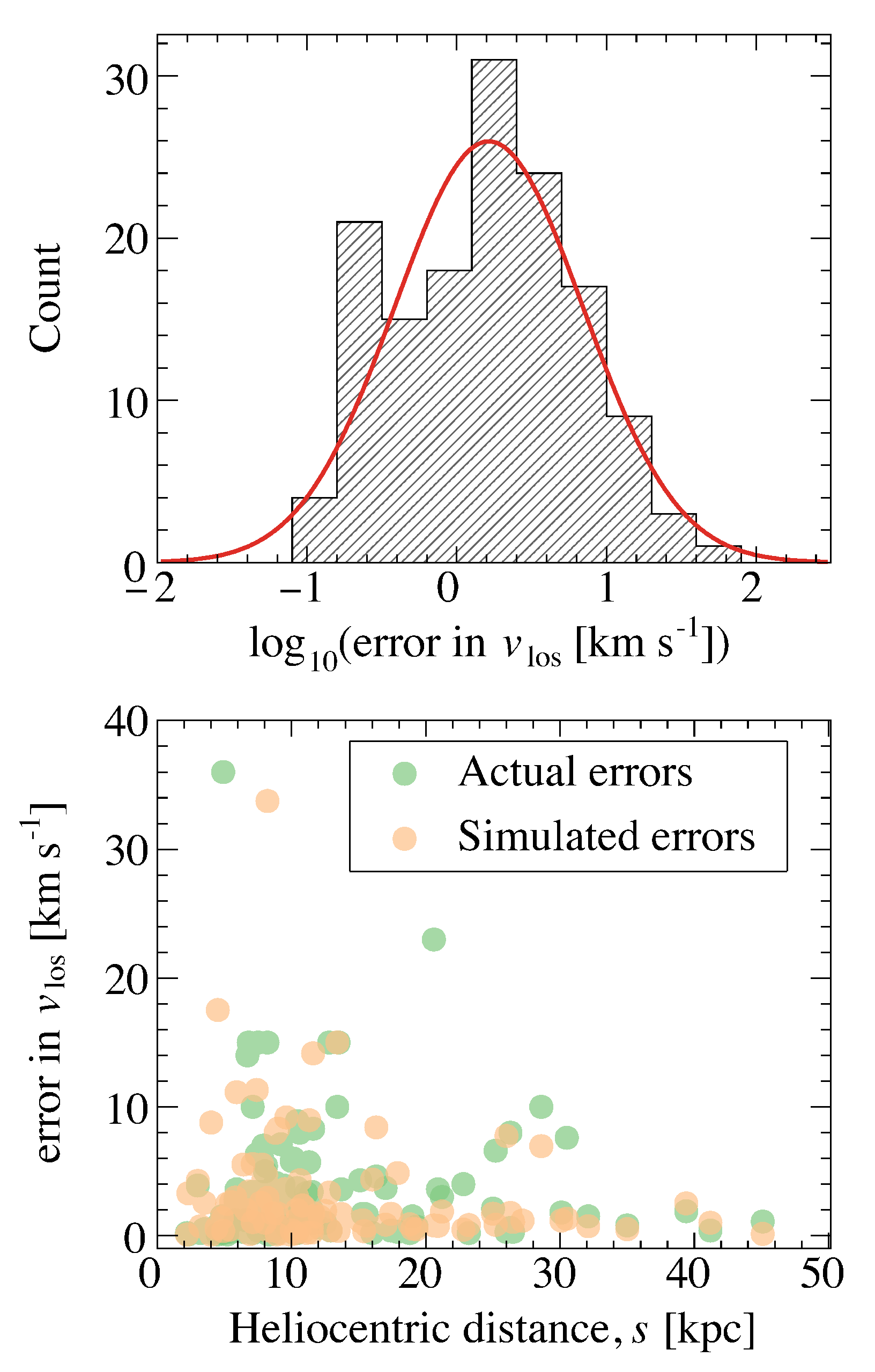}
\caption{(Top:) The distribution of errors in the line-of-sight velocity $v_\los$ is fitted to a lognormal distribution. The logarithm of the error has a mean of $0.444$ and a standard deviation of $1.38$. (Bottom:) Scatter plot of the error in $v_\los$ against the heliocentric distance $s$ of each GC in the Harris catalogue. Superimposed is a second scatter plot where the error is replaced by a randomly drawn value from the lognormal distribution.}
\label{fig:ErrorModel}
\end{figure}

\begin{figure}
\centerline{\includegraphics[width=.8\hsize]{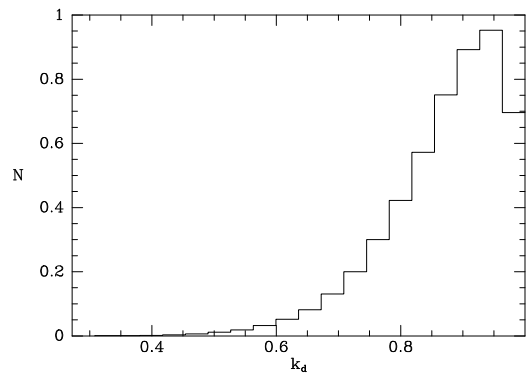}}
\caption{The posterior distribution of the parameter $k_\textup{d}$ when
the DF permits counter-rotating disc clusters.}\label{fig:kd}
\end{figure}

\begin{figure*}
\centering
\includegraphics[width=\textwidth]{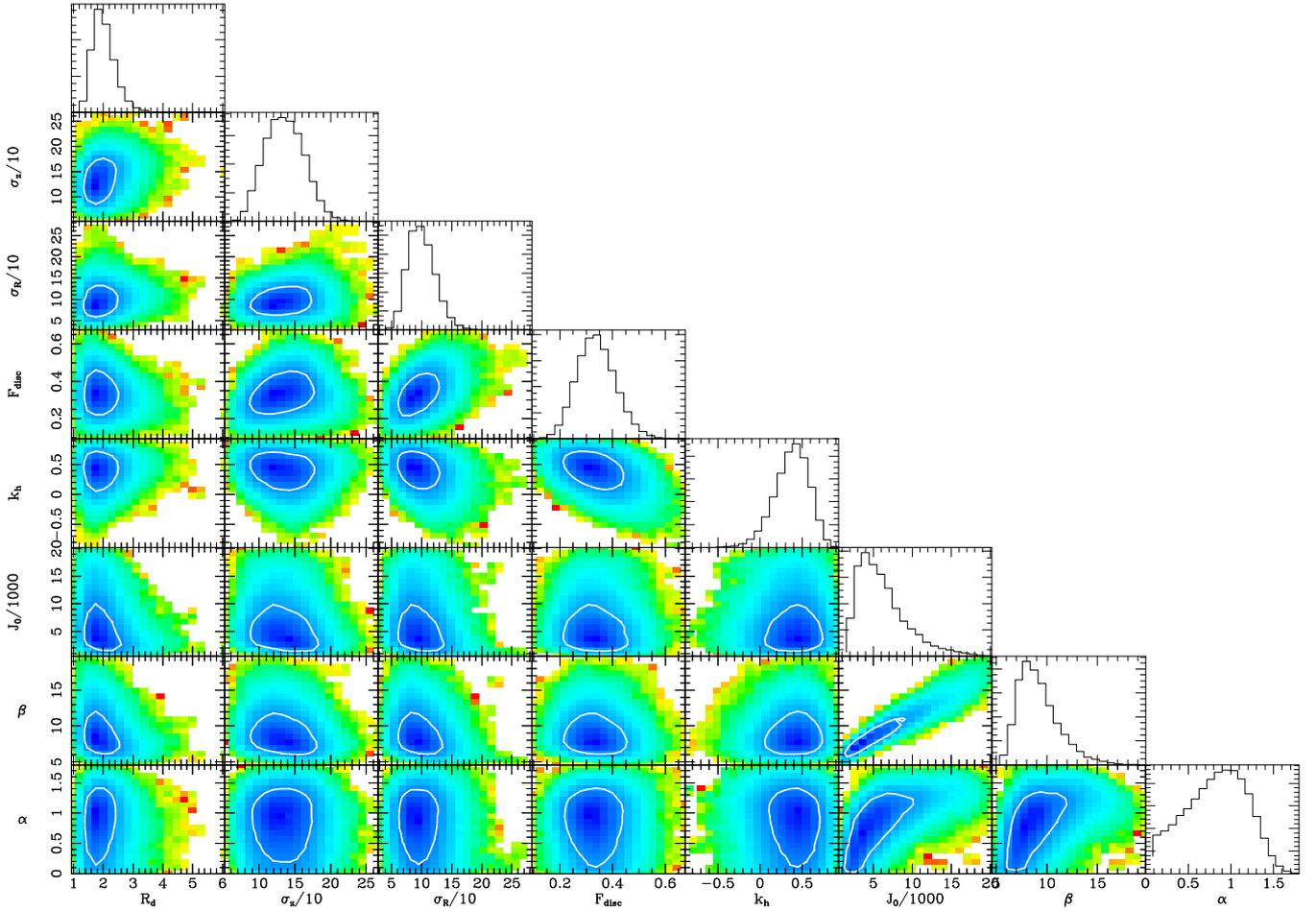}
\caption{The posterior probability distribution of models. The white contours
enclose 68 per cent of the probability.}\label{fig:correls}
\end{figure*}

\begin{figure}
\centerline{\includegraphics[width=.9\hsize]{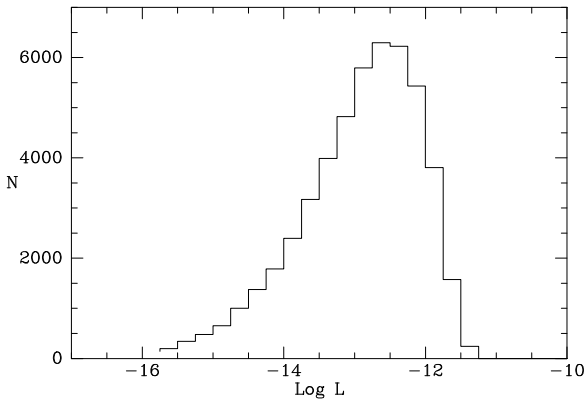}}
\caption{Distribution of the likelihoods $L$ of the  models sampled by
MCMC.}\label{fig:logL}
\end{figure}

\subsection{Posterior distribution}

We MCMC sample the posterior distribution as follows:
\begin{enumerate}
\item Sample $N_\textup{S}$ points from $f_\textup{S}(\bm x,\bm v)$ and
$N_\eta$ points for each GC from $\xi(\bm u|\bm u^\eta)$. These are
independent of $f(\bm J)$ so need only be sampled once at the beginning.

\item Use the \cite{SFD} map to determine the selection function $S(\bm u)$
for each of the above $N_\textup{S}$ points.

\item Use the St\"ackel fudge \citep{BinneyActions,BinneyIsochrones} with the
gravitational potential $\Phi$ in Section~\ref{sec:GalacticPotential} to
calculate $\bm J$ for each of these points.

\item Pick a point $\Pi$ in the space of model parameters at random.

\item\label{stepA} Calculate the prior $\textup{Pr}(\Pi)$ using
equations~\eqref{eq:PriorBeta} and \eqref{eq:Prior}.

\item If $\textup{Pr}(\Pi) \neq 0$, calculate the likelihood $\mathcal L$ for the model $f(\bm J|\Pi)$ using equations~\eqref{eq:McLikelihoodNorm} and \eqref{eq:McLikelihood}.

\item Calculate the posterior $\textup{Pr}(\Pi|\textup{Data})$ using equation~\eqref{eq:Posterior}.

\item\label{stepB} We use a robust adaptive Metropolis algorithm \citep{MCMC} to decide on the next point $\Pi$ in the Markov chain.

\item Repeat steps \ref{stepA} to \ref{stepB} until the desired number of MCMC points have been sampled.
\end{enumerate}
We have sampled $50\,000$ MCMC points, and we have used
$N_\eta = 100$ and $N_\textup{S} = 157 \times N_\eta$, the latter choice made
such that an equal number of points are evaluated for the numerator and
denominator of $\mathcal L$.

\subsection{Pseudo-catalogues}

We can probe the impact of noise, and the extent to which a model can match
the data
by using a model to generate
pseudo-catalogues of clusters. We generate a pseudo-catalogue as follows:
\begin{enumerate}
\item Sample a phase-space point $(\bm x,\bm v)$ for a `pseudo-GC' from
$f(\bm J(\bm x,\bm v)|\Pi)$ and compute  the
observables $\bm u$.

\item Accept the pseudo-GC with probability given by the selection function
$S(\bm u)$ and return to the previous step until 157 pseudo-GCs have been
accepted.

\item The observables of each pseudo-GC are ascribed errors
$\bm\sigma$ according to the error model described below.

\item The observables $\bm u$ are scattered by their errors.

\end{enumerate}

\def\pmin#1 #2 #3 {$#1^{+#2}_{-#3}$}
\begin{table}
\caption{Expectation values and standard deviations of the parameters of the
models in the MCMC chain.}\label{table:MCMCparams}
\begin{center}
\begin{tabular}{lcc}
\hline
Parameter & expectation & s.d. \\
\hline
$\alpha$ &     0.77 &     0.36 \\
$\beta$ &     8.83 &     2.01 \\
$J_0/\kpc\kms$ &     5650 &     3140 \\
$k_{\rm h}$ &     0.33 &     0.20 \\
$F_\textup{disc}$ &     0.32 &     0.07 \\
$\sigma_r/\kms$ &     94.3 &     22.5 \\
$\sigma_z/\kms$ &    130.3 &     26.7 \\
$\cR_{\rm d}/\kpc$ &     1.90 &     0.39 \\
\hline
\end{tabular}
\end{center}
\end{table}

The line-of-sight velocities $v_\los$ have errors that approximately follow a
lognormal distribution (see Fig.~\ref{fig:ErrorModel}). \emph{A priori}, we
might expect that these errors depend on the heliocentric distance $s$, but
by plotting errors in $v_\los$ versus distance $s$ in the Harris catalogue,
the lower plot of Fig.~\ref{fig:ErrorModel} demonstrates that errors in
$v_\los$ do not increase significantly with $s$. Consequently, each GC in a
pseudo-catalogue is given an error randomly drawn from the same lognormal
distribution.

As more than half of the GCs do not have proper motion data, quantities
calculated using the proper motions will come with very large uncertainties,
and we do not consider these useful to decide if a model agrees with data. We
therefore have not concerned ourselves with modelling the error distribution
of the proper motions, and will restrict ourselves to analysing
pseudo-catalogues using only the observables $(l,b,s,v_\los)$.

\section{Results}
\label{sec:Results}

\subsection{Parameters to fix}

Equation (\ref{eq:defk}) applied to the halo and disc DFs defines four
parameters $k_\textup{h}$, $L_\textup{h}$, $k_\textup{d}$ and $L_\textup{d}$
that control the amplitude and central steepness of each system's rotation
curve.  Along early MCMC chains the posterior distributions of $L_\textup{h}$
and $L_\textup{d}$ were indistinguishable from their priors (uniform in $\log
L_i$). Moreover, no correlation was apparent between $L_i$ and any other
parameter of the DF.  Evidently, the data do not usefully constrain the
steepness of the central rotation curves of the components, so in the
following we simply adopt $L_i=100\kpc\kms$, which we shall find causes
the rotation curves of the disc and halo populations to have central slopes
$\sim 100\kms\kpc^{-1}$.

Fig.~\ref{fig:kd} shows the posterior distribution of $k_\textup{d}$, which
is crowded near the upper limit of its permitted range.  The crowding of the
posterior distribution of $k_\textup{d}$ near unity implies a lack of
evidence for a counter-rotating disc of clusters. Consequently, in the
following we set $k_\textup{d}=1$.

In early MCMC exploration of parameter space we adopted a uniform prior on
the parameter $\gamma$ that controls the radial variation of the disc's
velocity dispersions (eqn.~\ref{eq:defs}).  The posterior distribution of
$\gamma$ then extended both sides of zero.  Since we consider increases in
velocity dispersion with radius implausible, we then ran chains with the
prior on $\gamma$ taken to be uniform in $\log\gamma$. The resulting
posterior distribution of $\gamma$ was essentially uniform in $\log\gamma$
for $\gamma<10^{-6}\kpc^{-1}$. That is, the data only require that $\gamma$
is so small as to have negligible impact in the region $r<100\kpc$ to which
the data are confined. This being so, we subsequently set $\gamma=0$, i.e.,
we made the velocity-dispersion parameters independent of $J_\phi$.

In summary, we fix the values of four parameters:
\begin{equation}
L_\textup{d}=L_\textup{h}=100\kpc\kms,\quad
k_\textup{d}=1,\quad \gamma=0,
\end{equation}
leaving the posterior distributions of eight parameters to be explored by
MCMC chains. Of these four describe the halo DF ($\alpha$, $\beta$,
$k_\textup{h}$ and $J_0$), three describe the disc DF ($\cR_\textup{d}$,
$\sigma_r$ and $\sigma_z$), and  the eighth and final adjustable parameter is
$F_\textup{disc}$, which is the fraction of the probability associated with
the disc.

\subsection{Favoured models}

Fig.~\ref{fig:correls} illustrates the structure of the posterior
distributions  of all adjustable parameters by showing on the diagonal
histograms for each parameter after marginalising over all other parameters,
and in the off-diagonal panels the correlations between each pair of model
parameters, again after marginalising over all other parameters. The white
contours in the off-diagonal panels enclose 68 per cent of the probability. 

\begin{figure*}
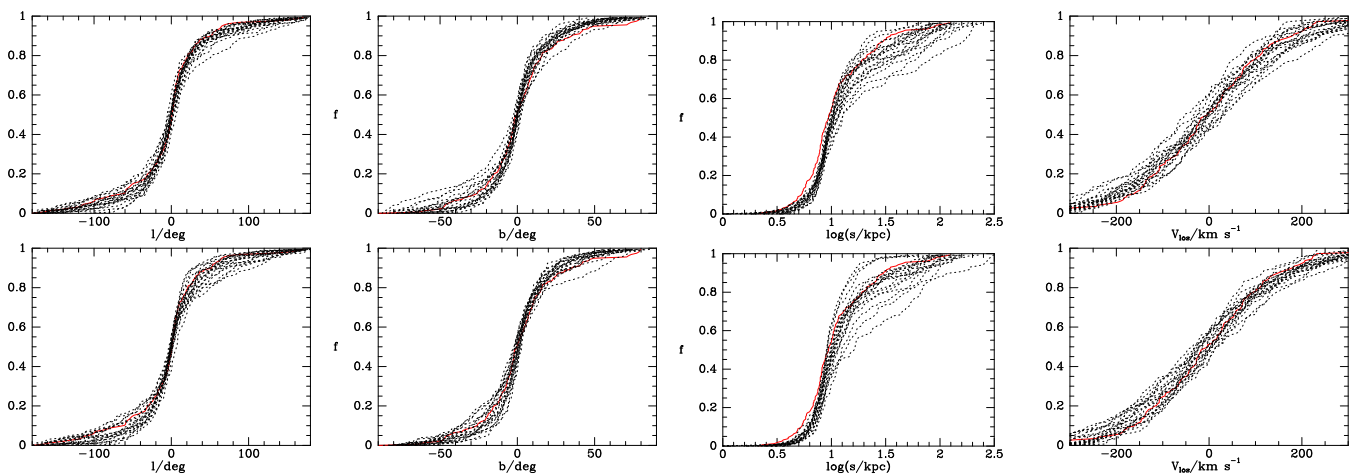

\centerline{\includegraphics[width=.24\hsize]{figs/hist0.ps}\quad
\includegraphics[width=.24\hsize]{figs/hist1.ps}\quad
\includegraphics[width=.24\hsize]{figs/hist2.ps}\quad
\includegraphics[width=.24\hsize]{figs/hist3.ps}}
\centerline{\includegraphics[width=.24\hsize]{figs/rand_hist0.ps}\quad
\includegraphics[width=.24\hsize]{figs/rand_hist1.ps}\quad
\includegraphics[width=.24\hsize]{figs/rand_hist2.ps}\quad
\includegraphics[width=.24\hsize]{figs/rand_hist3.ps}}
 \caption{Cumulative distributions of the observables, in red from the Harris
catalogue, and in black from 20 samples of 157 clusters drawn from (a) the
maximum-likelihood model (top row) and (b) 20 models randomly chosen from the
MCMC chain. All 157 clusters contribute to the first three panels in each
row, whereas only 143 clusters contribute to the end panels.}
\label{fig:hist}
\end{figure*}

\begin{figure}
\centerline{\includegraphics[width=.8\hsize]{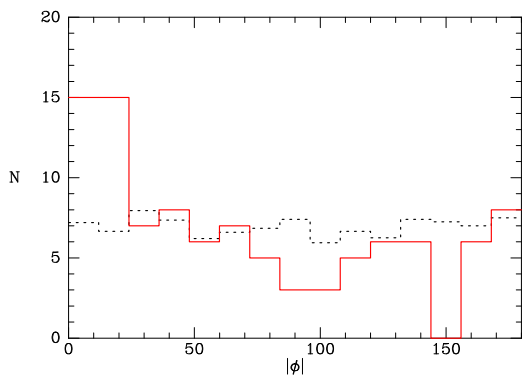}}
\caption{Full red histogram: the distribution of the galactocentric azimuths of the
clusters in the Harris catalogue with the Sun located at $\phi=0$. Dashed black histogram: the same for 20
catalogues of 157 clusters drawn from 20 models in the MCMC
chain.}\label{fig:phi}
\end{figure}

Parameters that are quite well determined are the extent to which the halo
cluster distribution rotates $k_{\rm h}$, the fraction of disc clusters $F_\textup{disc}$,
the velocity-dispersion parameters of the disc clusters $\sigma_r$ and
$\sigma_z$, and the scale length of the system of disc clusters $\cR_\textup{d}$.
For each parameter, Table~\ref{table:MCMCparams} gives  the expectation value
and the standard deviation along the MCMC chain.

Three parameters for the halo clusters are strongly correlated: the inner and
outer slope parameters $\alpha,\beta$ and the scale action $J_0$. Large
values of $J_0$ are associated with large values of $\alpha$ and $\beta$.
This correlation makes perfect sense physically when one recalls that
$\alpha$ and $\beta$ control the inner and outer slopes of the real-space
density profile of the population of halo clusters, while $J_0$ controls the
break radius that divides the two regimes.  In so far as the density profile
of the halo cluster distribution steepens smoothly with increasing radius,
such a correlation between $\alpha$, $\beta$ and $J_0$ and is inevitable.
Naively, one expects $J_0$ to be roughly the
product of the circular speed and the radius $\sim 2.5\kpc$ of the
break in the density profile, so $\sim600\kpc\kms$. The probable values of
$J_0$ are significantly larger than this. However, we show below
that the recovered parameter values do reproduce the expected break radius.

\subsection{What do we learn from the MCMC sample?}\label{sec:critique}

Fig.~\ref{fig:logL} shows the distribution of the likelihoods of the data
given the models sampled by MCMC. This distribution -- essentially the
$\chi^2$ distribution of the models -- is more than two orders of
magnitude wide, so significant probability is associated with models that
make the data hundreds of times less probable than does the most probable model.
This state of affairs is commonplace when models with significant numbers of
parameters are employed in Bayesian inference. 

Why must we consider models that make the data so much less
probable than the maximum-likelihood model? Because the
maximum-likelihood model achieves its
high likelihood in large measure by fitting not only the signal but also the
the noise in the data. In Fig.~\ref{fig:hist} we demonstrate this by
comparing the observables predicted by samples of 157 clusters drawn from (a)
the maximum-likelihood model (top row) with (b) models drawn at random from
the MCMC chain (lower row). We see that the scatter around the real
observables (red curves) of the observables predicted by the
maximum-likelihood model and the models drawn at random are indeed similar.
This result confirms that the excess likelihood of the maximum likelihood
model over typical models in the MCMC chain indeed reflects its ability to
fit the high level of noise inherent in there being only 157 clusters.

In Fig.~\ref{fig:hist} the only panels in which the red line of the data lies
outside the region explored by the 20 realisations are those for the
distance, $s$: in these panels the full red line for the data rises steeply
with increasing $s$ at values of $s$ that are $\sim10$ per cent smaller than
the broken curves of the pseudo-data. The number of clusters increases
rapidly as $s$ becomes comparable to $R_0$ and the clusters gathered around
the Galactic Centre enter the sample.

\begin{figure*}
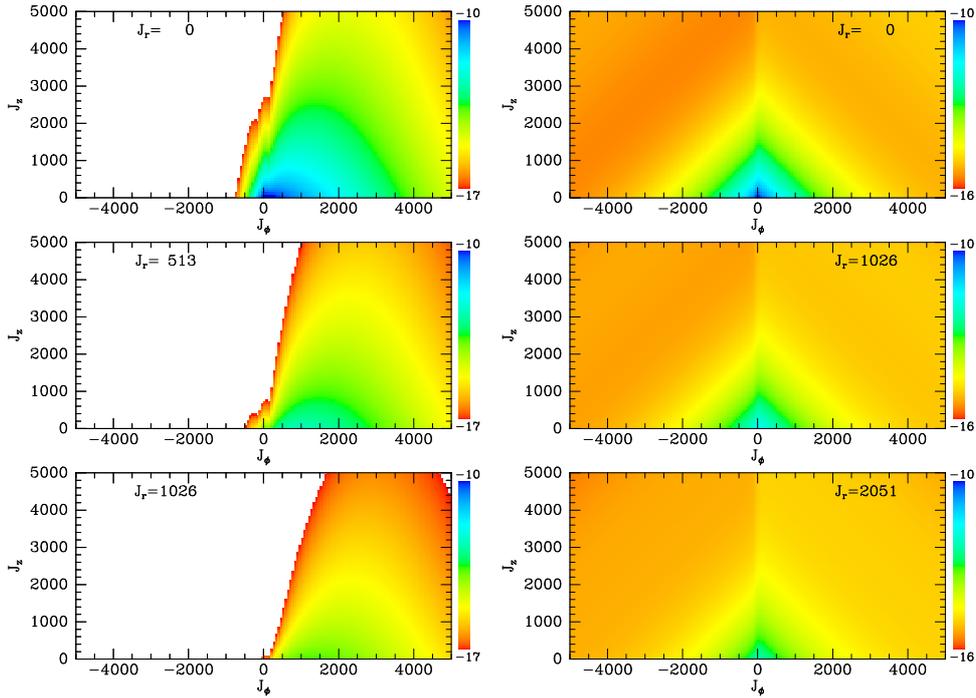

\centerline{
\includegraphics[width=.35\hsize]{figs/MultiSlicesD0.ps}\quad
\includegraphics[width=.35\hsize]{figs/MultiSlicesH0.ps}
}
\centerline{
\includegraphics[width=.35\hsize]{figs/MultiSlicesD4.ps}\quad
\includegraphics[width=.35\hsize]{figs/MultiSlicesH8.ps}
}
\centerline{
\includegraphics[width=.35\hsize]{figs/MultiSlicesD8.ps}\quad
\includegraphics[width=.35\hsize]{figs/MultiSlicesH16.ps}
}
 \caption{Left column: the phase-space probability density of disc clusters on three
 slices through phase space. Each slice is at the value of $J_r$ given in
 $\!\kpc\kms$ at top. Right column: the probability density of halo clusters on similar
 slices. The colour scale gives the log to base 10 of the phase-space
 density.
} \label{fig:Model}
\end{figure*}
\begin{figure}
\centerline{\includegraphics[width=.9\hsize]{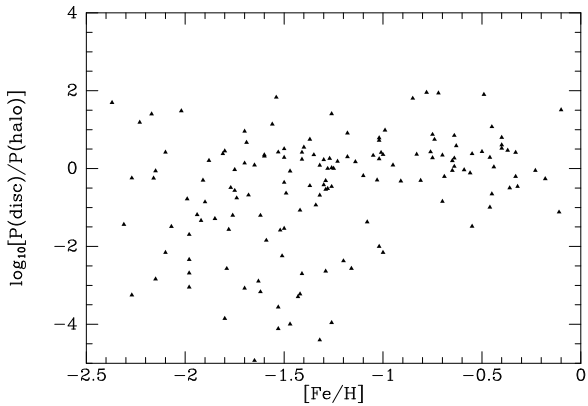}}
\caption{Clusters with $\hbox{[Fe/H]}\ga-1.4$ are assigned relatively high
probabilities by a typical disc DF. We plot vertically the ratio of the mean
(over MCMC sampled models) for each  object to be a disc or halo cluster.}\label{fig:dischalo}
\end{figure}

The red histogram in Fig.~\ref{fig:phi} shows the distribution of the
Galactocentric azimuths of the real clusters: there is a clear excess of
clusters with $|\phi|<20$ degrees, indicating that they lie in front of the
Galactic Centre. By contrast, the black dashed histogram, which shows the
azimuthal distribution of the clusters in twenty models from the MCMC chain,
is very uniform, so we have not succeeded in modelling the distribution of
clusters in the central region. The cause of this discrepancy could be either
(a) that distances to clusters are systematically too short, or (b) that our
value $R_0=8.3\kpc$ is too long, or (c) that observers have failed to
identify significant numbers of clusters that lie in or behind the bulge. The
latter could be due either to more extinction than we have adopted, or
confusion in crowded fields. However, following the {\it Vista
Variables in the Via Lactea} (VVV) survey \citep{Minnit2010} of the bulge
region in the near IR and with excellent spatial resolution, it seems
unlikely that item (c) is a significant problem.

Apart  from this question surrounding the distribution of cluster in the bulge
region, we consider that the plots of Fig.~\ref{fig:hist} are consistent with
the actual clusters being drawn from a model in the MCMC chain.

\subsection{The expected distribution of clusters}

An MCMC chain encodes the probability density of each part of model space,
and the DF quantifies the probability density of GCs in phase space for a
given model. Consequently, if we average $f({\bm x},{\bm v})$ over models in
an MCMC chain, we obtain our best estimate of the probability of finding a GC
at $({\bm x},{\bm v})$. We now present plots obtained by averaging $f$ over
50 models drawn from an MCMC chain.

Fig.~\ref{fig:Model} shows the probability
density of disc and halo clusters in action space. Specifically, the
logarithm to base 10 of the density is shown on three slices at constant
radial action, from $J_r=0$ (circular orbits) at the top to larger values of
$J_r$ lower down. The two types of cluster are seen to have very different
phase-space distributions. The disc clusters are most dense along the
$J_\phi$ axis and their density declines steeply with increasing $J_r$ or
$J_z$, whereas the halo clusters are dense only at the origin of action space
but their density declines relatively slowly with increasing $|{\bf J}|$. 

\begin{figure*}
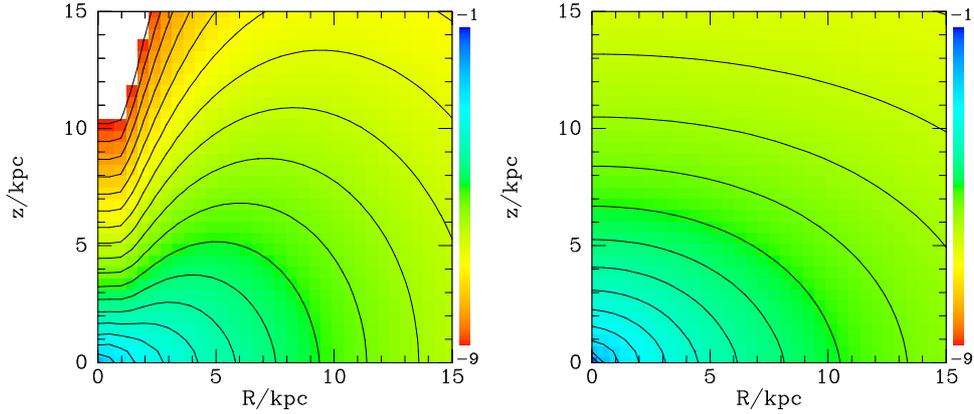

\centerline{\includegraphics[width=.35\hsize]{figs/plotMultiConts1.ps}\quad
\includegraphics[width=.35\hsize]{figs/plotMultiConts0.ps}}
\caption{The real-space density of disc clusters (left) and halo clusters
(right) on a plane that includes the symmetry axis. The colour scale shows the
log to base 10 of the density.}\label{fig:RZdensity}
\end{figure*}

The model provides probabilities for a cluster to be a disc rather than a
halo cluster based solely on the cluster's phase-space position. It is
natural to ask how these probabilities relate to the cluster's metallicities:
the latter have a clearly bimodal distribution with a minimum at
$\hbox{[Fe/H]}\simeq-0.8$ \citep{Zinn1985,Harris2016} dividing the metal-poor halo
cluster from the metal-rich disc clusters. 

Fig.~\ref{fig:dischalo} shows for each cluster in the Harris catalogue the
ratio $\langle P_{\rm disc}\rangle/\langle P_{\rm halo}\rangle$ versus
[Fe/H], where, for example $P_{\rm disc}$ is the integral in one model of the
disc DF over the cluster's error ellipsoid (eqn~\ref{eq:Likelihood}). For 73
clusters this ratio exceeds unity, so they are deemed more likely to be disc
than halo clusters.  Clusters with $\hbox{[Fe/H]}\ga-1.3$ are never
considered much more likely to be halo clusters than disc cluster, while only
a few clusters with $\hbox{[Fe/H]}<-2$ are much more likely to be disc than
halo clusters.  The only clusters that will be considered much more likely to
be halo than disc clusters are those that counter-rotate or lie at very large
radii.  Inevitably, at least as many halo clusters at moderate $r$ co- rather
than counter-rotate, and these clusters will have significant probabilities
to be disc clusters. Hence it is to be expected that more clusters have high
probabilities to belong to the disc than have $\hbox{[Fe/H]}>-0.8$. Hence
Fig.~\ref{fig:dischalo} is fully consistent with the conjecture that all
metal-rich clusters belong to the disc population, while the metal-poor
clusters all belong to the halo population. 

Fig.~\ref{fig:RZdensity} shows the real-space
density of disc clusters (left) and halo clusters (right) estimated from 50
models drawn at random from the MCMC chain. The 
colour scale shows the logarithm of the density in a slice that includes the $z$
axis.  For semi-major axes of length $a\la5\kpc$, the system of disc clusters has
the isodensity contours of an oblate body with axis ratio $\sim1:3$. Larger
isodensity surfaces have a deep depression around the minor axis. Any system
in which all particles rotate in the same sense about the $z$ axis can be
expected to have a low density on that axis.  The right panel of
Fig.~\ref{fig:RZdensity} indicates that the system of halo clusters forms a
simpler oblate structure of axis ratio $q\sim0.7$ out to semi-major axes
$a\sim20\kpc$, but a similar plot on a bigger scale shows that around
$r\sim30\kpc$ the system becomes quite spherical because the potential is
then dominated by the spherical dark halo rather than the disc.

\begin{figure}
\centerline{\includegraphics[width=\hsize]{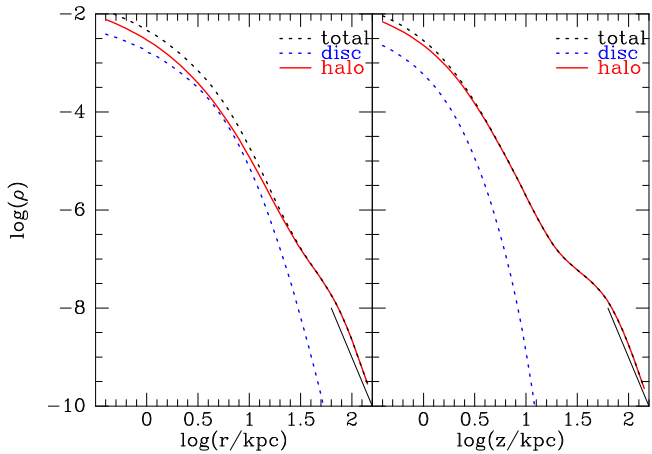}}
\caption{The probability densities of disc and halo
clusters as a function of radius in the equatorial plane (left) and along the
symmetry axis (right) estimated from 50 randomly chosen models in the MCMC
chain. The black straight line at the lower right of the left
panel has a slope of $-5$. }\label{fig:dens_plot}
\end{figure}

Fig.~\ref{fig:dens_plot} shows the density of clusters as a
function of radius in the equatorial plane (left panel) and along the $z$
axis (right panel). In each panel we show the contributions of the disc and
halo clusters (blue and red lines, respectively) and the total density in
black.  Straight black lines at the lower right of each panel have slope
$-5$, revealing that at the largest radii the cluster density is declining
somewhat faster than as $r^{-5}$. From results in \cite{Posti} it is easy to
show that if the rotation curve were flat, $\rho\sim r^{-5}$ would imply
$f\sim|\vJ|^{-5}$. Table~\ref{table:MCMCparams} indicates a markedly steeper decline in
the DF, $f\sim |\vJ|^{-9}$, because the MCMC chain favours large values of
the scale action $J_0$, so the asymptotic regime has not been reached even
at $r\sim80\kpc$.

The profiles of the system of halo clusters show wiggles at $R,z\ga20\kpc$.
The wiggle on the $z$ axis is most pronounced.  This feature disappears when
the disc's potential is removed and the  dark halo's potential is enhanced in
compensation, so it arises from the response of the cluster system to the
fading of the disc's gravitational field. It coincides with the decrease in
flattening noted above in connection with Fig.~\ref{fig:RZdensity}.

\begin{figure}
\centerline{\includegraphics[width=.9\hsize]{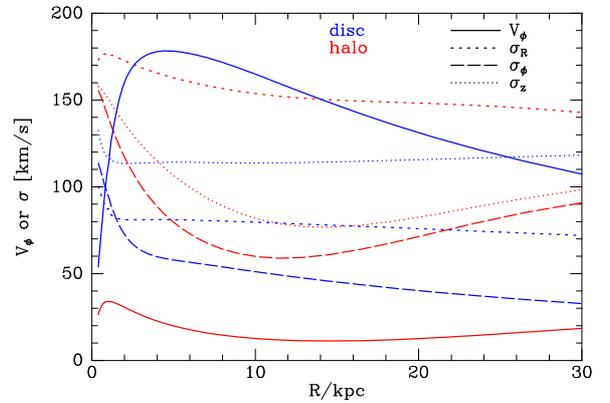}}
\caption{The kinematics of the disc (blue) and
halo (red) cluster systems estimated from 50 randomly chosen models in the
MCMC chain. The rotation curves show $\langle v_\phi\rangle$,
while various broken curves show the three principal dispersions, all in the
equatorial plane.}\label{fig:kin}
\end{figure}

Fig.~\ref{fig:kin} shows the kinematics computed from 50 models as a
function of radius in the equatorial plane. As the centre is approached, the
three broken red curves, which show $\sigma_R$, $\sigma_\phi$ and $\sigma_z$
in the halo system, approach one another at quite a large value:
$\sim170\kms$. While the centre of the halo system is isotropic, with
increasing $R$ the curves for $\sigma_\phi$ and $\sigma_z$ fall much more
steeply than the curve for $\sigma_R$, so the body of the system is quite
radially biased: the conventional anisotropy parameter $\beta_{\rm
s}$ reaches a peak value $0.68$ at $R=12\kpc$ and then gently falls to
$\beta_{\rm s}=0.53$ at $R=30\kpc$. The full red curve shows that the mean rotation rate of the
halo system is small -- it peaks at $\langle v_\phi\rangle\sim40\kms$ at
$R\sim1\kpc$ and from there falls to a minimum of $12\kms$ around $R=15\kpc$.

The disc system rotates fast: its rotation curve reaches $\langle
v_\phi\rangle=185\kms$ at $R=5\kms$ and from there $\langle v_\phi\rangle$
falls almost linearly to $118\kms$ at $R=30\kpc$. The velocity dispersion
tensor is quite anisotropic, with $\sigma_z\sim100\kpc$ being largest and
$\sigma_\phi$ falling from $\sim50\kms$ at $R=3\kpc$ to $\sim30\kms$ at
$R=30\kpc$.

As we reported in Section~\ref{sec:SamplingDensity}, the algorithm of
\cite{FrenkWhite} applied to the complete Harris catalogue yields a rotation velocity
$v_\textup{rot}=79\kms$. When we assemble a catalogue of 157
clusters from each of 20 models from the MCMC chain and apply the Frenk \&
White algorithm to each catalogue, the recovered values of $v_\textup{rot}$
have mean $69\kms$ and standard deviation $29\kms$. Thus our models are
consistent with the data from the perspective of rotation rate.

\section{Conclusions}
\label{sec:Conclusions}

We have constructed the first fully dynamical model of the Galaxy's GC system
that includes a realistic Galactic potential, that determined by
\cite{PifflRave}. The GC system is treated as a set of 157 identical,
non-interacting point particles in dynamic equilibrium orbiting in this
static potential. Motivated by the bimodal distribution of cluster
metallicities, the DF consists of two components: a disc and halo. After
fixing a number of parameters that were either barely constrained by the data
or essentially fixed by the data, the final model had eight parameters to be
fitted to the data: inner and outer slope parameters, a scale action and a
rotation rate for the halo, a scale length, in-plane and vertical velocity
dispersions for the disc, and the fraction of the probability provided by the
disc.

Although 157 clusters prove too few to constrain tightly any of the parameters, the
disc fraction, the three disc parameters and the halo's rotation rate all
produce well-defined peaks in the likelihood. The remaining three halo
parameters have a natural degeneracy, in which an increase in the scale
action (and corresponding scale radius) can be compensated by changes to the
inner and outer slopes of the density profile.

Given the similarities between the metallicity distribution functions of the
disc clusters and the thick disc \citep{GilmoreWyse1995}, the DF of the disc component
invites comparison with the Extended Distribution Function (EDF)
\citealt[][hereafter SB15]{SandersBinney} fitted to the stellar thick disc
using data from the Geneva-Copenhagen survey
\citep{Holmberg2007,Holmberg2009,CasagrandeGCS}.  We found that the cluster
data did not provide useful constraints on any radial variation in the
velocity-dispersion parameters $\sigma_r$ and $\sigma_z$ whereas SN15 found
that these decreased outwards with a scale length $\sim6.2\kpc$. If we
compare our values, $\sigma_r\sim94\kms$ $\sigma_z\sim130\kms$, with those,
$\sigma_r\sim\sigma_z\sim100\kms$, predicted by SB15 at $R\simeq4\kpc$, where
most of the disc cluster lie, the agreement is good except that we find
$\sigma_r<\sigma_z$. Our scale length $\cR_\textup{d}\sim1.9\kpc$ is a bit shorter
than that, $\cR_\textup{d}=2.3\kpc$ obtained by SB15, but entirely in line with
the findings of \citep{Bovy2016} from an analysis of APOGEE data
\cite{Majewski2016}. Even though we have used no metallicity information, our
value for the fraction of disc clusters, $F_{\rm disc}\sim0.32\pm0.07$, is
consistent with the $\sim31\pm9$ per cent of clusters that are metal-rich \citep{Harris2016}.

It is natural to compare our halo DF with the EDF fitted to halo K giants by
\cite{DasBinney}. For definiteness, we compare our metallicity-blind halo DF
with the EDF evaluated at $\hbox{[Fe/H]}=-2$.  The K giants required
essentially the same scale action ($\sim5000\kpc\kms$) as the clusters, but
our inner slope parameter ($\alpha\sim0.8$) is slightly smaller than that
($\alpha\sim1.3$) fitted to the K giants. Our outer slope parameter,
$\beta\sim8.8$, is definitely larger than that $\sim5$ fitted to the K
giants, with the consequence that in real space we predict the cluster
density falls off as $\sim r^{-5}$ rather than $\sim r^{-4}$ for the K
giants.  Given the degeneracy between $\alpha,\beta$ and $J_0$, it is not
clear that these differences are significant, but we have reason to expect we
might expect the distribution of GCs to be less centrally peaked than that of
halo giants: dynamical friction can drag globular clusters in to the densest
part of the bulge, where they will be tidally destroyed
\citep{Tremaine1975,Gnedin2014}. The axis ratio of the
distribution of K giants seems to increase from $q\sim0.7$ at small radii to
unity at large radii just as does that of the cluster distribution. 

We recover a probability density of clusters in real space that is consistent
with previous work \citep{BicaEtAl}. We note, however, that a DF such as our
halo DF, which is featureless in action space, gives rise to an interesting
feature at $r\sim10\kpc$ in real space as a consequence of the rapid decay of
the quadrupole in the disc's gravitational field. Specifically, as $r$
increases through $10\kpc$ the cluster system becomes spherical quite
rapidly. Since it is natural for the DF to have a simple form in action
space, a rapid reduction in flattening in all halo components around
$r\sim10\kpc$ is a robust prediction.

While neither the K giants \citep{DasBinney} nor Blue Horizontal Branch stars
show clear rotation \citep{DasBW}, the great majority of the halo DFs in our
MCMC chain have parts odd in $J_\phi$ that cause the system to rotate in the
same sense as the disc. The rate of rotation is, however, slow, typically
peaking at $\langle v_\phi\rangle<40\kms$, which is close to the upper limit
of the rotation of the system of K giants halo.  Samples of clusters drawn
from favoured models of the entire cluster system yield values of the measure
of rotation $v_{\rm rot}$ defined by \cite{FrenkWhite} that are consistent
with the value obtained from the real clusters.

The only respect in which mock catalogues extracted from favoured models
materially deviate from the data is the distribution of clusters close to the
Galactic centre. With our adopted distance to the Galactic centre,
$R_0=8.3\kpc$, the cluster distances in the Harris catalogue place
significantly more clusters in front of the Galactic centre than behind it.
Given that a small fractional error in the distance $s$ to a cluster near the
Galactic centre gives rise to a large change in the cluster's Galactocentric
distance $r$, uncertainty in the distribution of clusters around the
centre is inevitable. To make progress with this issue one would need to
reconsider the distance to every cluster within, say $r=5\kpc$ to ensure that
it is consistent with data that point to $R_0=8.3\kpc$. An alternative
explanation of the excess of clusters in front of the Galactic centre is that
we have under-estimated the bias arising from dust and confusion against
discovering GCs located behind the bulge.

In this paper we have developed a robust framework within which theories
about the GC system can be formulated and tested. With only 157 objects,
models cannot be tightly constrained, but as the currently very sparse
proper-motion data grow in volume and precision, it should be rewarding to
revisit the present analysis. In the next few years our knowledge of the
thick disc will increase markedly and this understanding should be
encapsulated in an EDF. In view of the preliminary results we have obtained
here, a promising line of enquiry would be to require the DF of the system of
disc clusters to coincide with that of the stellar thick disc and see what
halo DF is required to complement it.

As the data become more precise, it will be interesting to fit DFs separately to
the high- and low-metallicity clusters: will the parameters of the disc
and halo DFs that emerge from this exercise be similar to those found here?

Several interesting lines of enquiry are made possible by possession of a DF
for the GCs. For example, we expect halo GCs to be clustered in action space
as a consequence of more than one GC being stripped from a single satellite
galaxy. One could seek evidence for clustering of GCs by comparing the
frequency of pairs of objects at separation $\Delta$ in action space when (a)
both objects are real GCs, (b) one object is a real GC and one is a pseudo GC
drawn from the DF, and (c) both objects are pseudo GCs.  Another interesting
investigation would consider the speed with which a plausible GC DF would
evolve through the action of dynamical friction and tidal destruction since
the rates of both processes are fully specified by $\vJ$.

\section*{Acknowledgements}

We thank the referee and members of the Oxford dynamics group for valuable comments on
drafts of this paper.  The research leading to these results has received
funding from the European Research Council under the European Union's Seventh
Framework Programme (FP7/2007-2013)/ERC grant agreement no.\ 321067.




\bibliographystyle{mnras}
\bibliography{gcs}





\bsp	
\label{lastpage}
\end{document}